\documentclass[aps,preprint,showpacs,preprintnumbers,showkeys,eqsecnum,amsmath,amssymb]{revtex4}
\usepackage{graphics}
\usepackage{graphicx}
\usepackage{txfonts}
\usepackage{dcolumn}
\usepackage{subfigure}
\usepackage{mathrsfs}
\usepackage{bm}
\usepackage{amsmath,amssymb,epsfig}
\usepackage{float}

\begin{document}

\title{Holographic Superconductor/Insulator Transition with logarithmic electromagnetic field in Gauss-Bonnet gravity}

\author{ Jiliang {Jing}\footnote{Electronic address:
jljing@hunnu.edu.cn}}
\author{ Qiyuan Pan}
\author{ Songbai Chen}
\affiliation{ Institute of Physics and
Department of Physics, and Key Laboratory of Low Dimensional Quantum Structures and
Quantum Control of Ministry of Education, Hunan Normal University,
Changsha, Hunan 410081, P. R. China}

\vspace*{0.2cm}
\begin{abstract}

The behaviors of the holographic superconductors/insulator transition are studied by introducing a charged scalar field coupled with a logarithmic electromagnetic field in both the  Einstein-Gauss-Bonnet AdS black hole and soliton. For the Einstein-Gauss-Bonnet AdS black
hole, we find that: i) the larger coupling parameter of logarithmic
electrodynamic field $b$ makes it easier for the scalar hair to be
condensated; ii) the ratio of the gap frequency in conductivity
$\omega_g$ to the critical temperature $T_c$ depends on both  $b$ and the
Gauss-Bonnet constant $\alpha$; and iii) the critical
exponents are independent of the $b$ and $\alpha$. For the
Einstein-Gauss-Bonnet AdS Soliton, we show that the system is the
insulator phase when the chemical potential $\mu$ is small, but
there is a phase transition and the AdS soliton reaches the
superconductor (or superfluid) phase when $\mu$ larger than critical
chemical potential. A special property should be noted is that the
critical chemical potential is not changed by the coupling parameter
$b$ but depends on $\alpha$.

\end{abstract}

\pacs{11.25.Tq, 04.70.Bw, 74.20.-z, 97.60.Lf.}

\keywords{Holographic superconductors, logarithmic electromagnetic field, Einstein-Gauss-Bonnet AdS black hole, Einstein-Gauss-Bonnet AdS soliton}

\maketitle

\section{Introduction}

The anti-de Sitter/conformal field theory (AdS/CFT) correspondence \cite{Maldacena,polyakov,Witten} indicates that a weak coupling gravity theory in a $d$-dimensional anti-de Sitter spacetime can be related to a strong coupling conformal field theory on the $(d-1)$-dimensional boundary. The AdS/CFT duality is a powerful tool for investigating strongly coupled gauge theories, the application might offer new insight into the study of strongly interacting condensed matter systems where the perturbational methods are no longer available. Therefore, much attention has been given to the studies of the AdS/CFT duality to condensed matter physics and in particular to superconductivity
recently \cite{HartnollJHEP12, HorowitzPRD78,Nakano-Wen,Amado,
Koutsoumbas,Maeda79,Sonner,HartnollRev,HerzogRev,
Ammon:2008fc,Gubser:2009qm,CJ0,Gregory,Pan-Wang,
Ge-Wang,Brihaye,Gregory1009,Pan-Wang1,Cai-pGB}. In these studies   most of the holographically dual descriptions for a superconductor are based on a model that a simple Einstein-Maxwell theory coupled to a charged scalar.

Since Heisenberg and Euler \cite{euler} noted that quantum electrodynamics predicts that the electromagnetic field behaves nonlinearly through the presence of virtual charged particles,
the nonlinear electrodynamics has been an interesting subject for many years \cite{born,gibb,Olivera1,Hoffman-Gibbons-Rasheed,
Oliveira,HM,HMM,MC,Gurtug} because the nonlinear electrodynamics carries more information than the  Maxwell field. One of the important nonlinear electrodynamics is the logarithmic electromagnetic field which appears in the description of vacuum
polarization effects. The logarithmic terms were obtained as exact 1-loop corrections for electrons in a uniform electromagnetic field
background by Euler and Heisenberg \cite{euler}. Ref. \cite{HH} presented a Born-Infeld-like Lagrangian with a logarithmic term which can be added as a correction to the original Born-Infeld one. The logarithmic electromagnetic lagrangian takes the form $\mathcal{L_{BI}}=- b^2 \ln \left( 1+F^2/b^2\right)$ where $b$ is a coupling constant. The Lagrangian tends to the Maxwell case in the weak-coupling limit $b\rightarrow \infty $.

Within the framework of AdS/CFT correspondence, the properties of holographic superconductors for a given black hole or soliton depend on behavior of the nonlinear electromagnetic field coupled with the
charged scalar filed. We \cite{Jing2011} investigated the behaviors of the holographic superconductors by introducing a charged scalar field coupled with a Power-Maxwell field and found that the larger power parameter makes it harder for the scalar hair to be condensated. We \cite{Jing} also studied the holographic superconductors in Gauss-Bonnet gravity with  Born-Infeld electrodynamics and noted that the model parameters and the Born-Infeld coupling parameter will affect the formation of the scalar hair, the transition point of the phase transition from the second order to the first order, and the relation connecting the gap frequency in conductivity with the critical temperature.
In this paper we will study the behaviors of the holographic superconductors/insulator transition by introducing a charged scalar field coupled with a logarithmic electromagnetic field in both the  Einstein-Gauss-Bonnet AdS black hole and soliton, and to see how the logarithmic electromagnetic field affect the formation of the scalar hair, the critical exponent, and the critical chemical potential of the systems.

The paper is organized as follows. In Sec. II, we give the holographic dual of the  Einstein-Gauss-Bonnet AdS black hole by introducing a charged scalar field coupled with a logarithmic electromagnetic field. In Sec. III, the behaviors of the holographic superconductors/insulator transition are studied by coupling a charged scalar field with a logarithmic electromagnetic field in the  Einstein-Gauss-Bonnet AdS soliton. We summarize and discuss our conclusions in the
last section.

\section{Einstein-Gauss-Bonnet AdS$_5$ black hole and Superconductor}

The action of the Einstein-Gauss-Bonnet theory, which is the most general Lovelock theory in five and six dimensions, is given by
\begin{equation}
I_{grav}=\frac{1}{16\pi G}\int\limits_{\mathcal{M}}d^{d}x\,\sqrt{-g}\,\left[
R-2\Lambda +\hat{\alpha} \,\left( R^{2}-4R_{\mu \nu }R^{\mu \nu }+R_{\mu \nu
\lambda \sigma }R^{\mu \nu \lambda \sigma }\right) \right] \,,
\end{equation}%
where  $\Lambda=-(d-1)(d-2)/(2L^2) $ is the cosmological constant,
and $\hat{\alpha} $ is the
Gauss-Bonnet coupling constant. The static spacetime in the  Einstein-Gauss-Bonnet gravity is described by
\cite{Boulware-Deser,Cai-2002}
\begin{eqnarray}\label{BH metric}
ds^2=-f(r)dt^{2}+\frac{dr^2}{f(r)}+r^{2}dx_{i}dx^{i},
\end{eqnarray}
with
\begin{eqnarray}
f(r)=\frac{r^2}{2\alpha}\left[1-\sqrt{1-\frac{4\alpha}{L^{2}}
\left(1-\frac{ML^{2}}{r^{d-1}}\right)}~\right],
\end{eqnarray}
where the constant $M$ is relate
to the black hole horizon by $r_{+}=(ML^{2})^{1/(d-1)}$ and $\alpha=\hat{\alpha}(d-3)(d-4)$. We can define the effective asymptotic AdS scale as
$
L^2_{\rm eff}=2\alpha/(1-\sqrt{1-\frac{4\alpha}{L^2}})
$
because $f(r)\sim\frac{r^2}{2\alpha}\left(1-\sqrt{1-4\alpha/L^2} \right)$ in the asymptotic region ($r\rightarrow\infty$).
The Hawking temperature of the Einstein-Gauss-Bonnet AdS black hole,
which will be interpreted as the temperature of the CFT, can be expressed as
\begin{eqnarray}
\label{Hawking temperature} T=\frac{(d-1)r_{+}}{4\pi L^{2}}\ .
\end{eqnarray}

In the background of the $d$-dimensional Einstein-Gauss-Bonnet AdS black hole, we consider the logarithmic electrodynamic field and the charged scalar field coupled via a generalized action
\begin{eqnarray}\label{System}
S=\int d^{d}x\sqrt{-g}\left[
-2 b^2 \ln \Big(1+\frac{F_{\mu\nu}F^{\mu\nu}}{8 b^2}\Big)-\frac{1}{2}\partial_{\mu}\tilde{\psi}\partial^{\mu}\tilde{\psi}
-\frac{1}{2}m^2\tilde{\psi}^2-\frac{\tilde{\psi}^2}{2}(\partial_{\mu}p-A_{\mu})
(\partial^{\mu}p-A^{\mu})  \right] \ ,
\end{eqnarray}
where $F^{\mu\nu}$ is the strength of the logarithmic electrodynamic field $F=dA$, $\tilde{\psi}$ is a scalar field, and $b$ is a coupling constant.  The logarithmic electrodynamic field will reduce to the Maxwell case as $b\rightarrow \infty$. We can
use the gauge freedom to fix $p=0$ and take $\psi\equiv\tilde{\psi}$ and $A_{t}=\phi$, where both $\psi$ and $\phi$ are  real functions  of $r$ only. Then  we can obtain the following equations of motion
\begin{eqnarray}\label{Psi}
&&\psi^{\prime\prime}+\left(
\frac{f^\prime}{f}+\frac{d-2}{r}\right)\psi^\prime
+\frac{\phi^2}{f^2}\psi-\frac{m^2}{f}\psi=0\,,
\\ \label{Phi} &&
\phi^{\prime\prime} \left(1+\frac{\phi^{\prime 2}}{4 b^2}\right)
+\frac{d-2}{r}\left(1-\frac{\phi^{\prime 2}}{4 b^2}\right)\phi^\prime-\frac{2\psi^2\phi}{f}\left(1-\frac{\phi^{\prime 2}}{4 b^2}\right)^2=0,
\end{eqnarray}
where  a prime denotes the derivative with respect to $r$. At the
event horizon $r=r_+$ of the black hole, we must have
\begin{eqnarray}\label{Nearh}
 &&\phi(r_{+})=0,\nonumber \\
 &&\psi(r_{+})=-\frac{(d-1)r_+ }{ m^{2}L^2}\psi^\prime(r_{+}),
\end{eqnarray} and at the asymptotic AdS region
($r\rightarrow\infty$), the solutions behave like
\begin{eqnarray}
&&\psi=\frac{\psi_{-}}{r^{\lambda_{-}}}+
\frac{\psi_{+}}{r^{\lambda_{+}}}, ~~~~~
\phi=\mu-\frac{\rho}{r^{d-3}}, \label{infinity}
\end{eqnarray}
with
\begin{eqnarray}
&&\lambda_\pm=\frac{1}{2}\left[(d-1)\pm\sqrt{(d-1)^{2}
+4m^{2}L_{eff}^2}~\right], \label{LambdaZF}
\end{eqnarray}
where $\mu$ and $\rho$ are interpreted as the chemical potential and
charge density in the dual field theory, respectively. The
coefficients $\psi_+$ and $\psi_-$ both multiply normalizable modes
of the scalar field equations and they correspond to the vacuum
expectation values $\psi_+=\langle \mathcal{O_+}\rangle$,
$\psi_-=\langle \mathcal{O_-}\rangle$ of an operator $\mathcal{O}$
dual to the scalar field according to the AdS/CFT correspondence. We
can impose boundary conditions that either $\psi_+$ or $\psi_-$
vanishes in the following study.

\subsection{Relations between critical temperature and charge density}

In this subsection we will present a detail analysis of the condensation of the operator $\langle{\cal O}_{+}\rangle$ and  $\langle{\cal O}_{-}\rangle$ by taking numerical integration of  the equations (\ref{Psi}) and (\ref{Phi})    from the event horizon out to the infinity with the boundary  conditions mentioned above.

\begin{figure}[htp!]
\includegraphics[scale=1.0]{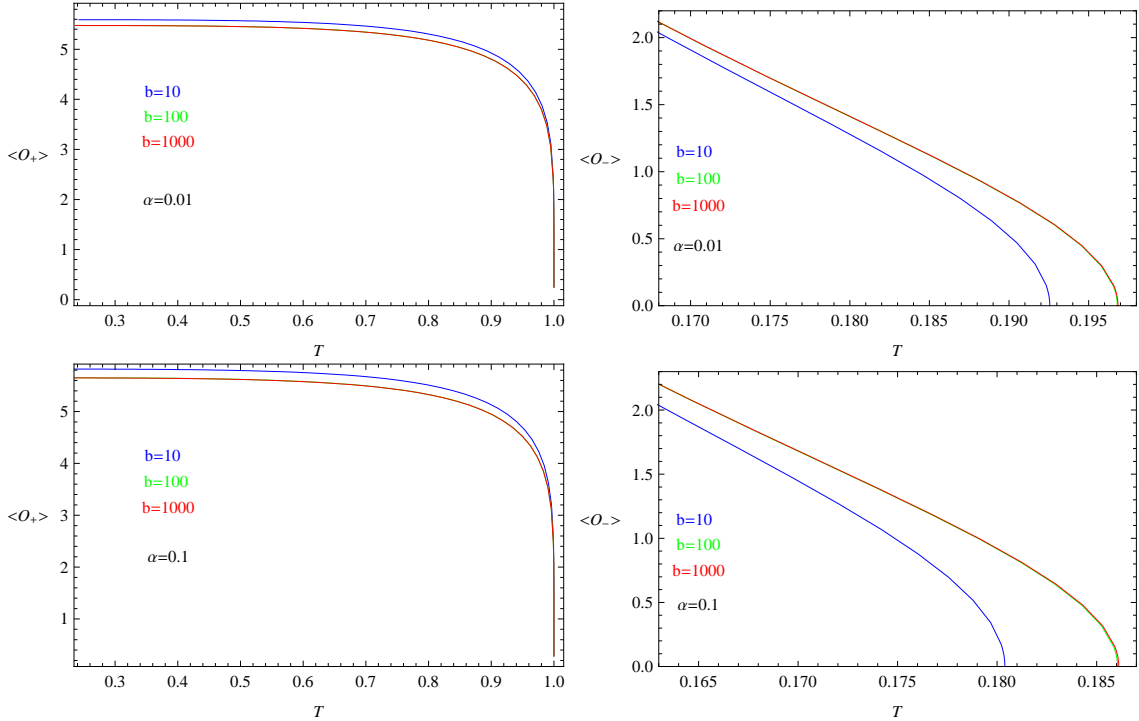}\hspace{0.2cm}%
\caption{\label{Fig5}  (Color online.) The condensate  for operators
$\langle{\cal O}_{+}\rangle$ (left column) and $\langle{\cal
O}_{-}\rangle$ (right column) as a function of the temperature for
different values of the coupling parameter $b$ and  Gauss-Bonnet
parameter $\alpha$. In each panel the lines for $100$ (green) and
$1000$ (red) are almost overlap.}\label{condensate}
\end{figure}

The influence of the parameters  $b$ and $\alpha$ on the condensation with fixed values $ m^2 L_{eff}^2=-3$ is presented in Fig. \ref{condensate}, in which two panels in the left column for the operator $\langle{\cal O}_{+}\rangle$ and panels in the right column are for the operator $\langle{\cal O}_{-}\rangle$.
We know from the figure that the value of the operator $\langle{\cal O}_{+}\rangle$ decreases as the coupling parameter $b$ of the logarithmic electrodynamics increases with a fixed  Gauss-Bonnet parameter $\alpha$, but the value of the operator $\langle{\cal O}_{-}\rangle$ increases as $b$ increases for the same $\alpha$. It should point out that the curves for both $b=100$ and $b=1000$ almost overlap.

In Table \ref{Tc-a}, we list the values of the critical  temperature
for different values of $b$ and $\alpha$ in the 5-dimensional black
hole, respectively. For both the scalar operators
$\langle\mathcal{O}_+ \rangle$ and $\langle\mathcal{O}_- \rangle$,
we find from the table  \ref{Tc-a} that the critical temperature
increases as the value of $b$ increases with fixed $\alpha$, which
means that the larger parameter $b$ makes it easier for the scalar
hair to be condensated in 5-dimensional Einstein-Gauss-Bonnet AdS
black hole; however,  the critical temperature decreases as $\alpha$
increases for the fixed $b$, which means that the stronger
Gauss-Bonnet coupling makes  condensate harder in the black hole.
\begin{table}
\begin{center}
\caption{\label{Tc-a} The critical values of $T_c$ for different $b$ and $\alpha$ in the 5-dimensional Einstein-Gauss-Bonnet AdS black hole with $m^2L_{eff}^2=-3.75$.}
\begin{tabular}{c || c | c || c | c || c | c }   \hline
 & \multicolumn{2}{c||}{$ b=1000 $} &\multicolumn{2}{c||} {$ b=100 $} &\multicolumn{2}{c}{$ b=10 $} \\
         \hline
 & $ T_c$  for $\langle\mathcal{O}_+ \rangle$ & $ T_c$  for $\langle\mathcal{O}_- \rangle$&
$ T_c$ for $\langle\mathcal{O}_+ \rangle$& $ T_c$  for $\langle\mathcal{O}_- \rangle$ &  $ T_c$ for $\langle\mathcal{O}_+ \rangle$ &$ T_c$  for $\langle\mathcal{O}_- \rangle$ \\
         \hline
$ \alpha=0.00$ &$0.219643 \rho^{1/3}$ &$0.312928 \rho^{1/3}$ & $0.219618 \rho^{1/3}$  &
$0.312925 \rho^{1/3}$& $0.217230\rho^{1/3}$ &$0.312580\rho^{1/3}$
                 \\
                   \hline
$ \alpha=0.01$ &$0.219004 \rho^{1/3}$ &$0.312078 \rho^{1/3}$ & $0.218979 \rho^{1/3}$  &
$0.312074 \rho^{1/3}$& $0.216555\rho^{1/3}$ &$0.311725\rho^{1/3}$
                 \\
\hline $ \alpha=0.05$ &$0.216291 \rho^{1/3}$ & $0.308494\rho^{1/3}$& $0.216264 \rho^{1/3}$
& $0.308491\rho^{1/3}$& $0.213676\rho^{1/3}$ &$0.308119\rho^{1/3}$
                 \\
\hline $ \alpha=0.10$ &$0.212456 \rho^{1/3}$ & $0.303548\rho^{1/3}$& $0.212436
\rho^{1/3}$ & $0.303543  \rho^{1/3}$ &$0.209578 \rho^{1/3}$ & $0.303139
\rho^{1/3}$
        \\
        \hline
\end{tabular}
\end{center}
\end{table}

\subsection{Critical exponents}

\begin{figure}[htp!]
\includegraphics[scale=1.]{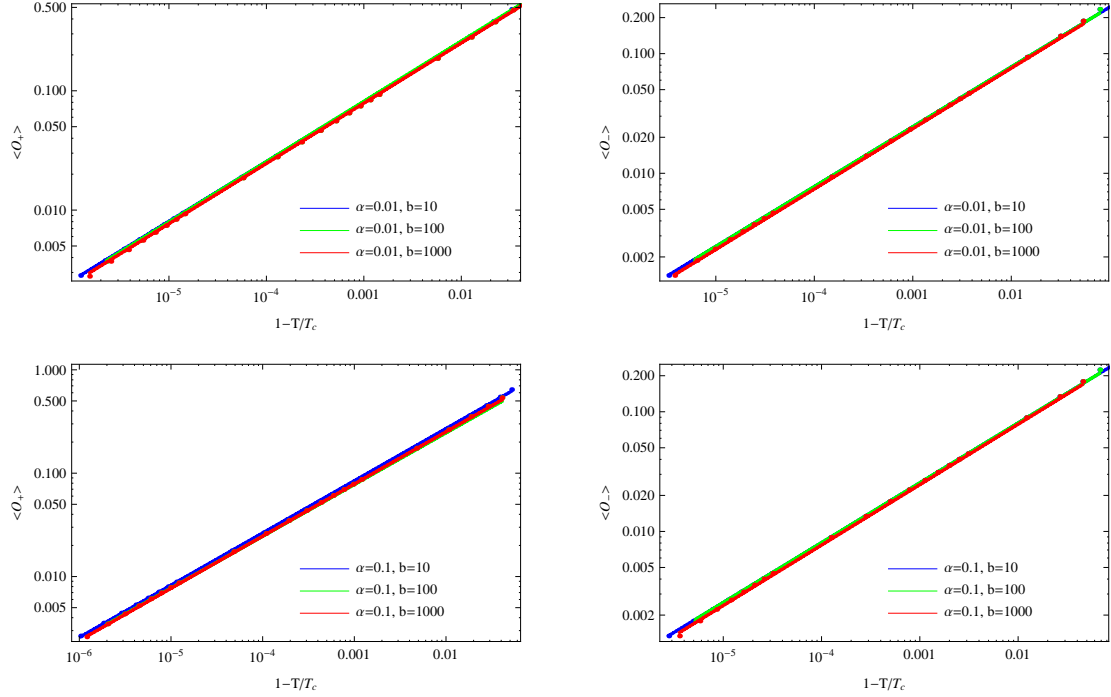}\hspace{0.2cm}
\caption{\label{Fig1}  (Color online.) The critical exponents for
operators $\langle{\cal O}_{+}\rangle$ (left column) and
$\langle{\cal O}_{-}\rangle$ (right column) as a function of the
temperature for different values of the coupling parameter $b$ and
Gauss-Bonnet parameter $\alpha$. In each panel the three lines for
$b=10$ (blue), $100$ (green) and $1000$ (red) are almost overlap.}
\label{CE}
\end{figure}

In this subsection, we will study the critical exponents of the
holographic superconductor model with the logarithmic
electromagnetic field by using numerical method. In Fig. \ref{CE},
we present the condensate of the operators $\langle\mathcal{O}_+
\rangle$ (left column) and $\langle\mathcal{O}_- \rangle$ (right
column) as a function of $(1-T/T_{c})$ in logarithmic scale with
different values of $b$ and $\alpha$ for $d=5$. The top two panels
for $\alpha=0.01$ and bottom two for $\alpha=0.1$, and in each panel
the three lines  correspond to $b=10$, $100$  and $1000$ which are
almost overlap. We see  from these panels that the slope is
independent of the parameters $b$ and $\alpha$, which is in
agreement with the value $1/2$. It should be pointed out that the result seems to be a universal
property for various nonlinear electrodynamics if the scalar field
$\psi$ takes the form of this paper.

\subsection{Electrical Conductivity }

Horowitz {\it et al.} \cite{HorowitzPRD78} got following relation connecting the gap frequency in conductivity with the critical temperature for (2+1) and (3+1)-dimensional superconductors
\begin{eqnarray}
\frac{\omega_g}{T_c}\approx 8,
\end{eqnarray}
which is roughly twice the BCS value 3.5 indicating that the holographic superconductors are strongly coupled. We now examine this relation for the Gauss-Bonnet gravity with the logarithmic electrodynamic field.

In the computation of the electrical conductivity, the perturbation of the logarithmic electromagnetic field should be considered in the  Gauss-Bonnet black hole background. However, in the probe approximation we can ignore the effect of the perturbation of the metric. Assuming that the perturbation is translational symmetric and has a time dependence as $\delta A_x=A_x(r)e^{-i\omega  t}$, we find that the motion equation for the logarithmic electrodynamic field in the Gauss-Bonnet black hole background can be written as
\begin{eqnarray}
&&\left[A_{x}^{\prime\prime}+\left(\frac{f^\prime}{f}
+\frac{d-4}{r}\right)A_{x}^{\prime}+\frac{\omega^2}{f^2}A_{x}\right]
\left(1-\frac{\phi^{\prime 2}}{4b^2}\right)\nonumber \\
&&+\frac{\phi^{\prime }\phi^{\prime \prime}}{2b^2}A_{x}^{\prime}
-\left(1-\frac{\phi^{\prime 2}}{4b^2}\right)^2\frac{2\psi^2}{f}A_{x}=0 \; .
\label{Maxwell Equation}
\end{eqnarray}
Noting that an ingoing wave  boundary condition near the horizon is
$
A_{x}(r)\sim f(r)^{-\frac{i \omega L^2}{(d-1) r_+}}, $  and a general behavior for $d=5$ in the asymptotic AdS region ($r\rightarrow\infty$) \cite{Gregory1009} is
$
A_{x}=L_e^{-1/2} A^{(0)}+\frac{L_{eff}^{3/2}}{r^2}\left(A^{(2)}-\frac{1}{2} \ln \frac{r}{L} \partial_t^2 A^{(0)}  \right),
$
we find that the holographic conductivity can be expressed as  \cite{Gregory1009}
\begin{eqnarray}\label{GBConductivity}
\sigma=\frac{2A^{(2)}}{i\omega A^{(0)}}+\frac{i\omega}{2} -i \omega \log\frac{L_e}{L} \ ,
\end{eqnarray}
Solving the motion equation (\ref{Maxwell Equation}) numerically we can obtain the conductivity.
\begin{figure}[htp!]
\includegraphics[scale=1.0]{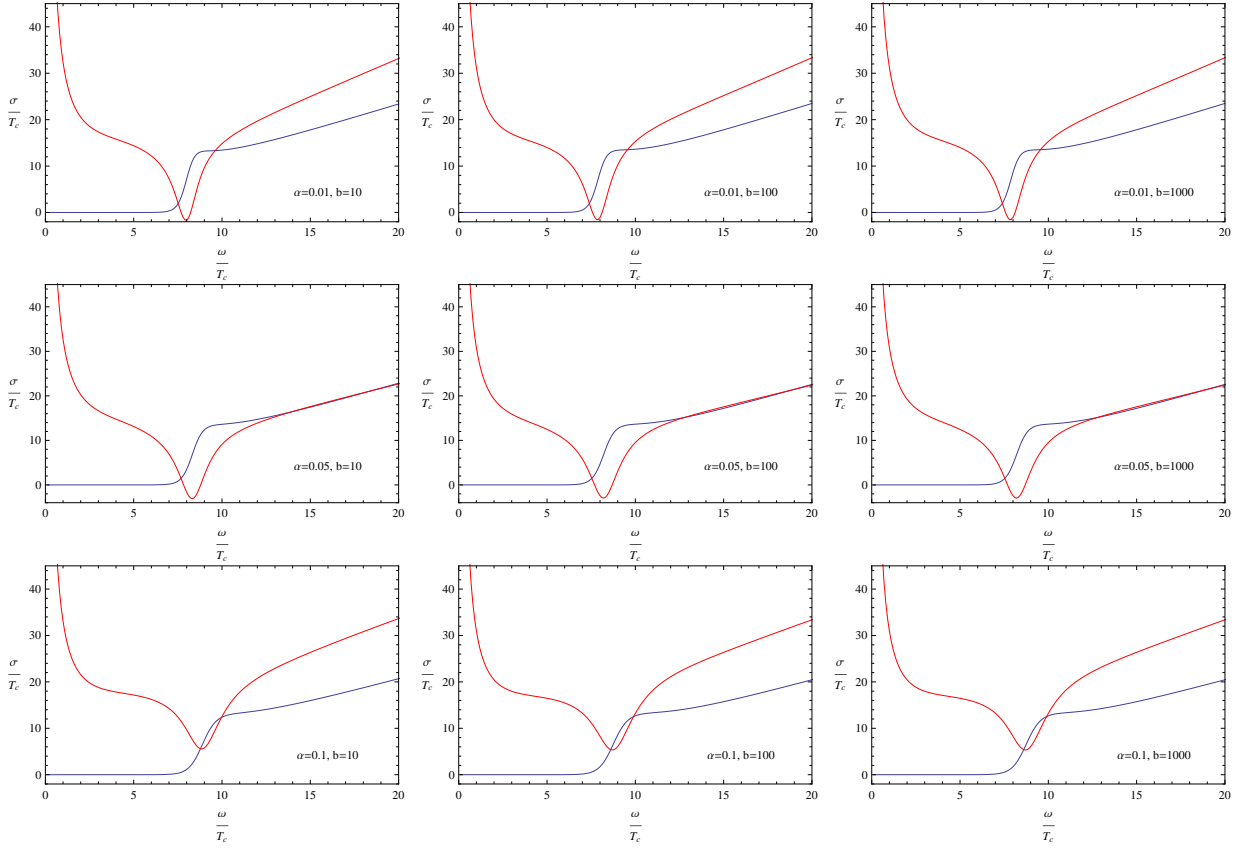}\hspace{0.2cm}%
\caption{\label{Fig6} (Color online) The conductivity of the superconductors as a function of $\omega/T_c$ for different values of $b$ and $\alpha$ with fixed $m^2L_{eff}^2=-3$. In each panel the blue (bottom)  line represents the real part of the conductivity, $Re(\sigma)$, and red (top) line is the imaginary part of the conductivity, $Im(\sigma)$.}\label{cond}
\end{figure}

\begin{table}
\caption{\label{ConductivityTc} The ratio $\omega_{g}/T_{c}$ for
different values of the Gauss-Bonnet constant $\alpha$ and the coupling parameter $b$ with $m^{2}L_{eff}^2=-3$.}
\begin{tabular}{|c|c|c|c|}
         \hline
~ &~~b=1000~~&~~b=100~~&~~b=10~
          \\
        \hline
~~$\alpha=0.01~~$~~ & ~~$7.7$~~ & ~~$7.8$~~ & ~~$7.9$~~
          \\
        \hline
~~$\alpha=0.05~~$ & $ 8.0 $ & $8.1 $ & $8.2 $
          \\
        \hline
~~$\alpha=0.10$~~ & $8.7$ & $8.8$ & $8.9$
\\
\hline
\end{tabular}
\end{table}

We present in Fig. \ref{cond} and table \ref{ConductivityTc} the frequency dependent conductivity obtained by solving the motion equation of the logarithmic electrodynamic field  numerically for different values of $\alpha$ and  $b$ with  $m^{2}L_{eff}^{2}=-3$  (we plot the conductivity at temperature  $T/T_c\simeq  0.25$).
We note that the gap frequency $\omega_{g}$ decreases with the increase of the coupling parameter $b$ for fixed  $\alpha$,  it increases as $\alpha$ increases for fixed $b$. From Fig. \ref{cond} and table  \ref{ConductivityTc}, we find that the ratio of the gap frequency in conductivity $\omega_g$ to the critical temperature $T_c$ in the Gauss-Bonnet black hole with the logarithmic electrodynamic field depends on both the Gauss-Bonnet constant and the coupling parameter of logarithmic electrodynamic field.

\section{Einstein-Gauss-Bonnet AdS soliton and Holographic  Superconductor}

In this section we will study a holographic superconductor  for a
logarithmic electrodynamic field coupled with a charged scalar field
in a Einstein-Gauss-Bonnet AdS soliton.
The metric of the Einstein-Gauss-Bonnet AdS soliton is  described by
\cite{Cai-Kim-Wang}
\begin{eqnarray}\label{soliton}
ds^2=-r^{2}dt^{2}+\frac{dr^2}{f(r)}+f(r)d\varphi^2 +
r^{2}h_{ij}dx^{i}dx^{j},
\end{eqnarray}
where $
f(r)=\frac{r^2}{2\alpha}\left[1-\sqrt{1-\frac{4\alpha}{L^{2}}
\left(1-\frac{r_s^{d-1}}{r^{d-1}}\right)}~\right].
$
This spacetime does not have any horizon but a conical  singularity at
$r=r_{s}$ which can be removed by imposing a period
$\beta=\frac{4\pi L^{2}}{(d-1)r_{s}}$ for the coordinate $\varphi$.

Using the generalized action (\ref{System}) for the  logarithmic
electrodynamic field coupled with the charged scalar field, we find that the
equations of motion are given by
\begin{eqnarray}
&&\psi^{\prime\prime}+\left(
\frac{f^\prime}{f}+\frac{d-2}{r}\right)\psi^\prime
+\left(\frac{\phi^2}{r^2f}-\frac{m^2}{f}\right)\psi=0\,,
\label{Psib}
\\
&&\left(1+\frac{ f \phi^{\prime 2}}{4 b^2 r^2}\right)
\phi^{\prime\prime} +\left(\frac{f^\prime}{f} +\frac{d-4}{r}\right)
\phi^\prime-\frac{(d-2)f}{4b^2r^3}\phi^{\prime 3}-\left(1-\frac{f
\phi^{\prime 2}}{4b^2 r^2}\right)\frac{2\psi^2}{f}\phi=0.
\label{Phib}
\end{eqnarray}
The asymptotic solutions of the functions $\psi$ and  $\phi$ near
the AdS boundary ($r\rightarrow\infty$) are the same as Eq.
(\ref{infinity}), but at the tip $r=r_{s}$ of the soliton, the
solutions must satisfy
\begin{eqnarray}
&&\psi=\tilde{\psi}_{0}+\tilde{\psi}_{1}\log(r-r_{s})
+\tilde{\psi}_{2}(r-r_{s})+\cdots\,, \nonumber \\
&&\phi=\tilde{\phi}_{0}+\tilde{\phi}_{1}\log(r-r_{s})
+\tilde{\phi}_{2}(r-r_{s})+\cdots\,, \label{SolitonBoundary}
\end{eqnarray}
where $\tilde{\psi}_{i}$ and $\tilde{\phi}_{i}$ ($i=0,1,2,\cdots$)
are the integration constants. We should impose  the Neumann-like
boundary condition $\tilde{\psi}_{1}=\tilde{\phi}_{1}=0$
\cite{Nishioka-Ryu-Takayanagi} to keep every physical quantity
finite. The probe approximation will also be used in our discussion.

\begin{figure}[htp!]
\includegraphics[scale=1.]{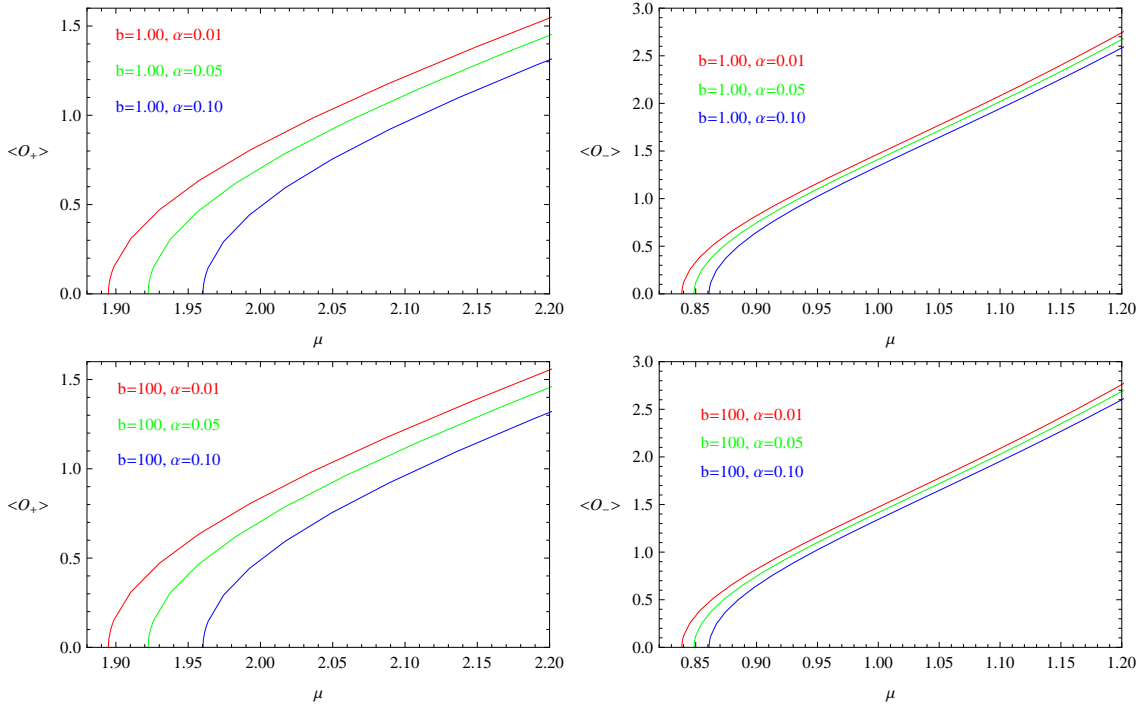}\hspace{0.2cm}%
\caption{\label{Cond-SolitonBH} (Color online) The condensates of
the scalar operators $\langle{\cal O_{+}}\rangle_S$ (left column) and $\langle{\cal O_{-}}\rangle_S$ (right column) with respect to the chemical potential $\mu$ in the Gauss-Bonnet Soliton.
From top row  to bottom one we take $b=1$ and $b=100$ with fixed $m^{2}L_{\rm eff}^2=-3.75$. In each panel, the three lines from left to right  correspond to
increasing $\alpha$, i.e., $\alpha=0.01$ (red), $0.05$ (green) and $0.1$ (blue) respectively.}
\end{figure}
\begin{figure}[htp!]
\includegraphics[scale=1.]{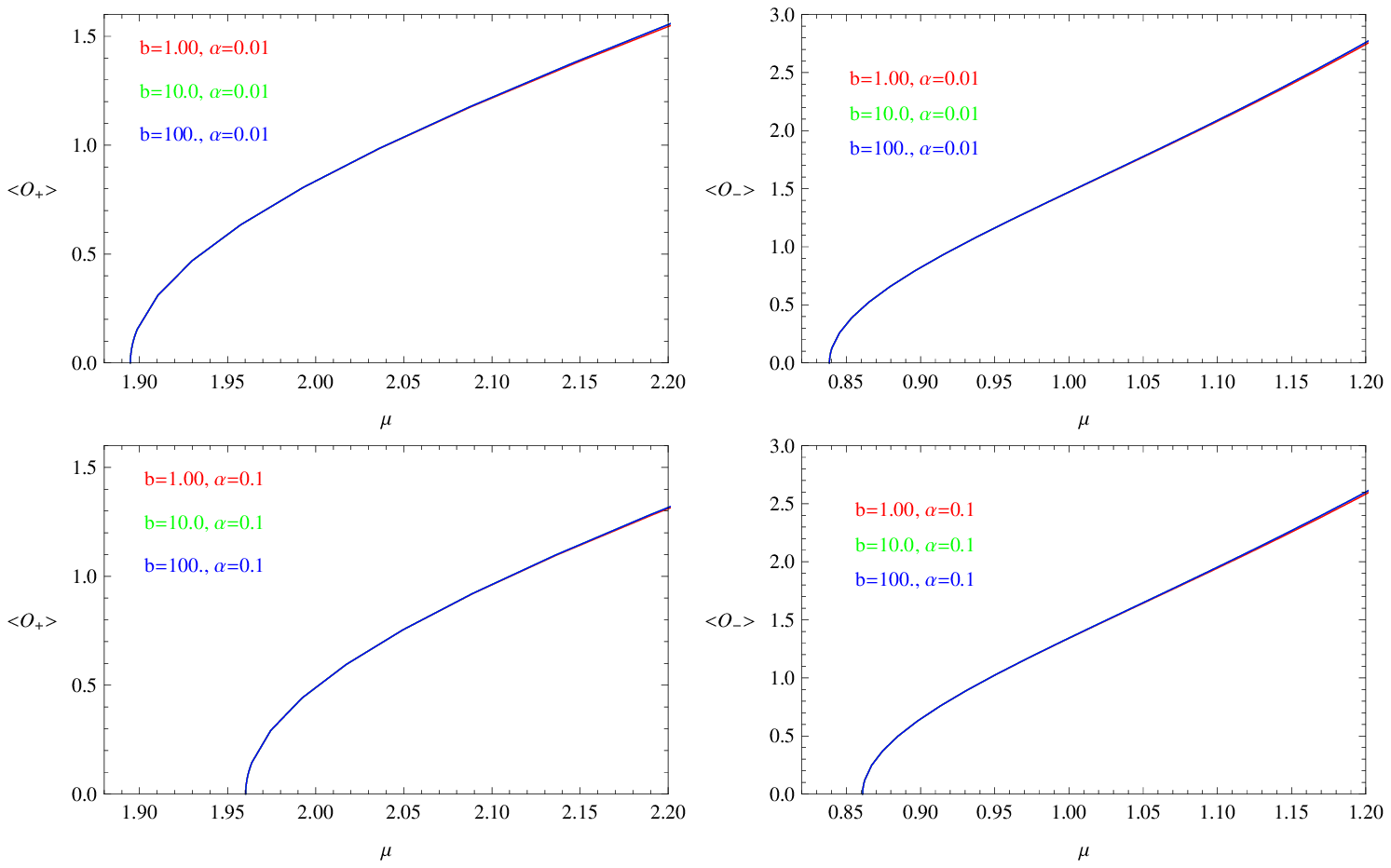}\hspace{0.2cm}%
\caption{\label{Fig3} (Color online) The condensates of the scalar
operators $\langle{\cal O_{+}}\rangle_S$  (left column) and
$\langle{\cal O_{-}}\rangle_S$ (right column) with respect to the
chemical potential $\mu$ in the Gauss-Bonnet Soliton. From top row
to bottom one we take $\alpha=0.01$ and $\alpha=0.1$ with fixed
$m^{2}L_{\rm eff}^2=-3.75$. In each panel the three lines for  $b=1$
(red), $10$ (green) and $100$ (blue) are almost overlap.}
\end{figure}

In Figs. \ref{Cond-SolitonBH} and \ref{Fig3} we plot the
condensations of scalar operators $\langle{\cal O_{+}}\rangle_{S}$
(left column) and $\langle{\cal O_{-}}\rangle_{S}$ (right column)
with respect to the chemical potential $\mu$ in the
Einstein-Gauss-Bonnet AdS Soliton for different logarithmic
electrodynamics parameter $b$ and Gauss-Bonnet coupling constant
$\alpha$ with the fixed scalar mass $m^{2}L_{\rm eff}^2=-3.75$. The
condensation occurs for scalar operators $\langle{\cal
O}_{i}\rangle_{S}$ ($i=\pm$) with different values of $b$ and
$\alpha$  if $\mu>\mu_{i S}$, where $\mu_{i S}$ is the so-called
critical chemical potential for scalar operators $\langle{\cal
O}_{i}\rangle_{S}$ which just begin to condense.  The $\mu_{+ S}$
and $\mu_{- S}$ for scalar operators $\langle{\cal
O_{+}}\rangle_{S}$ and $\langle{\cal O_{-}}\rangle_{S}$ with
different values of $b$ and $\alpha$ are listed in table \ref{Cond}.
From the figures and table we find that $\mu_{i S}$ increase as $\alpha$
increases with fixed $b$, while $\mu_{i S}$ does not change for different $b$ with fixed $\alpha$.

\begin{table}
\begin{center}
\caption{\label{Cond} The critical chemical potential $\mu_{+ S}$ and $\mu_{- S}$ for the scalar operators $\langle{\cal O_{+}}\rangle_{S}$ and $\langle{\cal O_{-}}\rangle_{S}$ in the 5-dimensional Einstein-Gauss-Bonnet AdS soliton with $m^2L_{eff}^2=-3.75$.}
\begin{tabular}{c || c | c || c | c || c | c }   \hline
 & \multicolumn{2}{c||}{$ b=1000 $} &\multicolumn{2}{c||} {$ b=100 $} &\multicolumn{2}{c}{$ b=10 $} \\
         \hline
 & $ \mu_{+ S}$ & $\mu_{- S}$&
$ \mu_{+ S} $& $\mu_{- S}$ &  $ \mu_{+ S}$ &$ \mu_{- S}$ \\
    \hline
$ \alpha=0.00$ &$1.88832 $ &$0.83618 $ &
$1.88832 $  & $0.83618 $& $1.88832 $
&$0.83618$
                 \\
         \hline
$ \alpha=0.01$ &$1.89484 $ &$0.83863 $ &
$1.89484 $  & $0.83863 $& $1.89484 $
&$0.83863$
                 \\
\hline $ \alpha=0.05$ &$1.92234 $ & $0.84860 $&
$1.92234 $ & $0.84860 $& $1.92234 $
&$0.84860 $
                 \\
\hline $ \alpha=0.10$ &$1.96025 $ & $0.86098 $&
$1.96025 $ & $0.86098  $ &$1.96025
$ & $0.86098 $
        \\
        \hline
\end{tabular}
\end{center}
\end{table}

\begin{figure}[htp!]
\includegraphics[scale=1.]{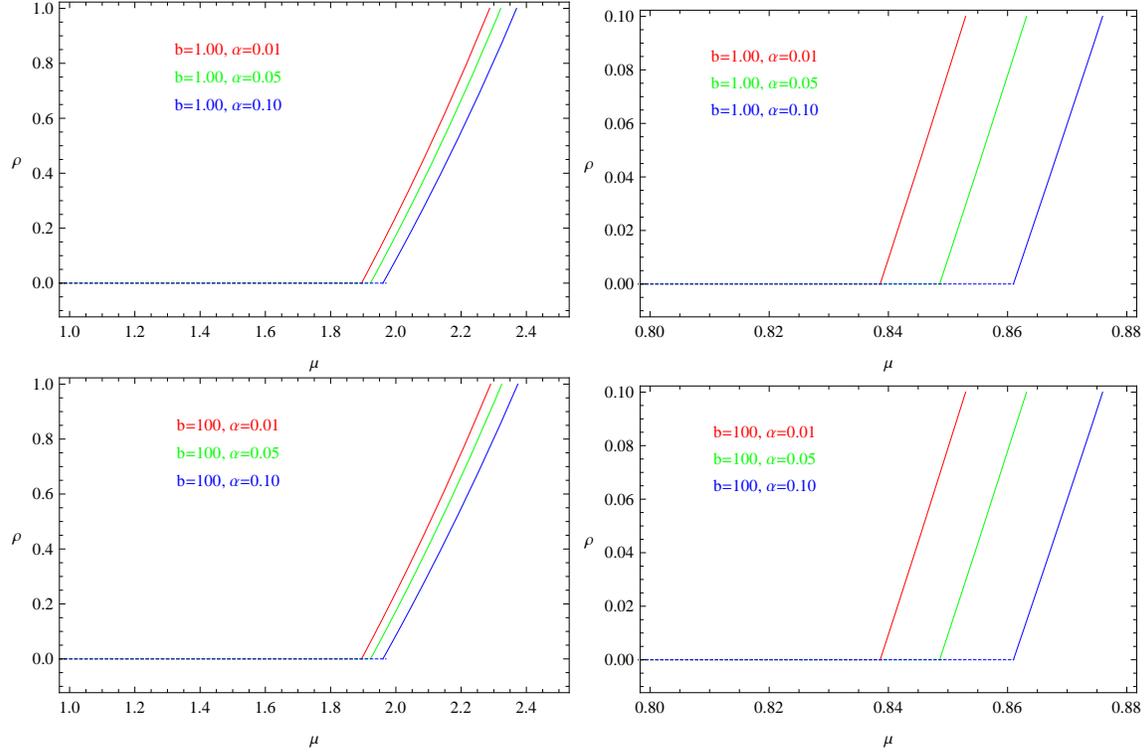}\hspace{0.2cm}%
\caption{\label{SLab2} (color online) The charge density
$\rho$ as a function of the chemical potential $\mu$ with different
values of $b$ and $\alpha$ when $\langle{\cal O_{+}}\rangle_S\neq0$
(left) and $\langle{\cal O_{-}}\rangle_S\neq0$ (right). From top row
to bottom one we take $b=1$ and $b=100$ with fixed
$m^{2}L_{\rm eff}^2=-15/4$. The three lines from left to right
correspond to increasing $\alpha$, i.e., $0.01$ (red), $0.05$
(green) and $0.1$ (blue) respectively.}
\end{figure}
\begin{figure}[htp!]
\includegraphics[scale=1.0]{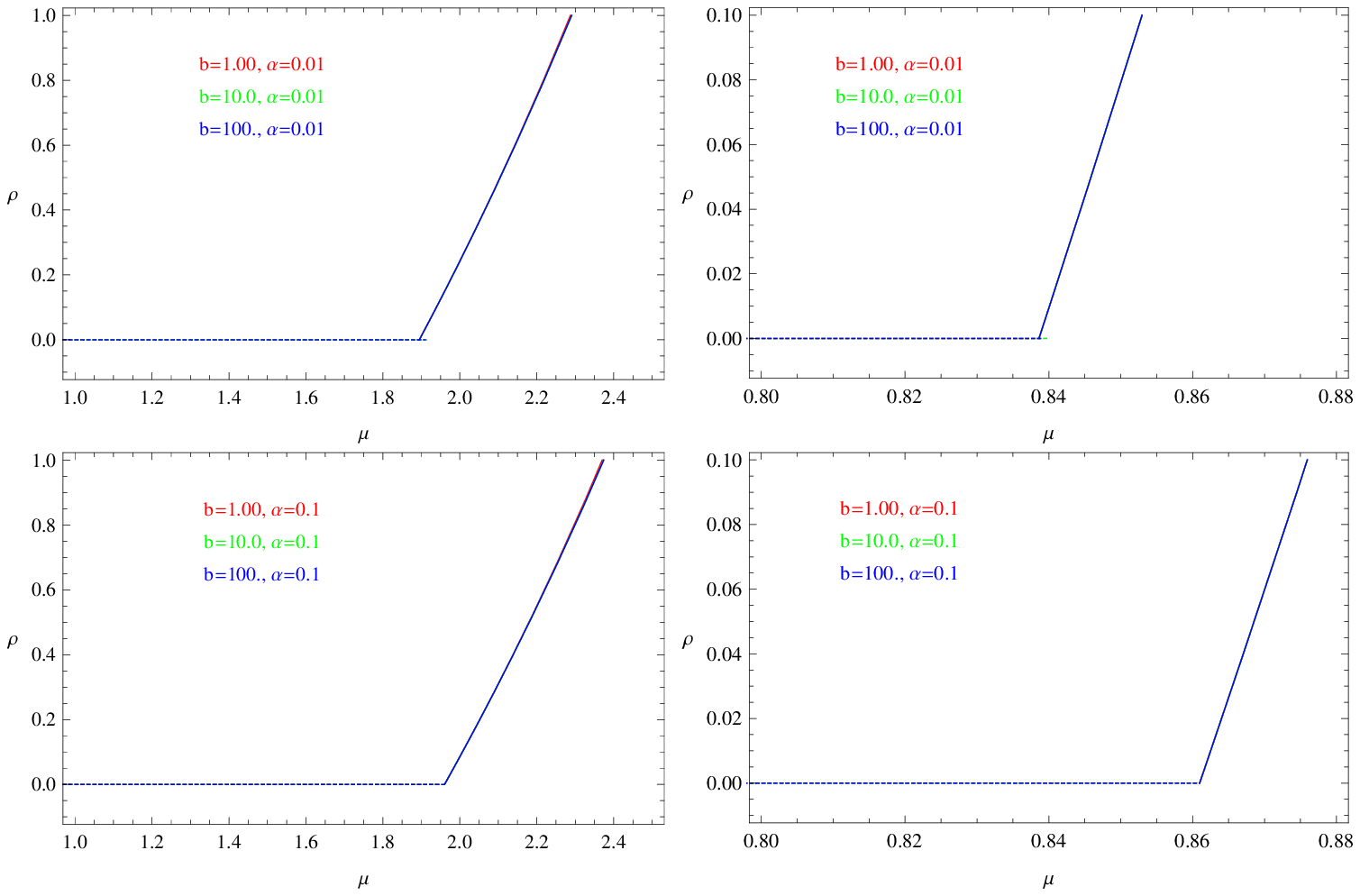}\hspace{0.2cm}%
\caption{\label{SLab4} (color online) The charge density
$\rho$ as a function of the chemical potential $\mu$ with different
values of $\alpha$ when $\langle{\cal O_{+}} \rangle_S\neq0$ (left)
and $\langle{\cal O_{-}}\rangle_S\neq0$ (right) with fixed mass
$m^{2}L_{\rm eff}^2=-3.75$. From top row to bottom one we take
$\alpha=0.01$ and $\alpha=0.1$. In each panel the three lines for
$b=1$ (red), $10$ (green) and $100$ (blue) are almost overlap.}
\end{figure}

Figs. \ref{SLab2} and \ref{SLab4} show that the charge density $\rho$ as a function of the chemical potential $\mu$ when $\langle{\cal
O_{+}}\rangle\neq0$ (left) and $\langle{\cal O_{-}}\rangle\neq0$ (right).
We see from these figures that, when $\mu$ is small, the system is described by the AdS soliton solution itself which is the insulator phase \cite{Nishioka-Ryu-Takayanagi}, however, when $\mu$ reaches $\mu_{+ S}$ or $\mu_{- S}$, there is a phase transition and the AdS soliton reaches the superconductor (or superfluid) phase. Fig. \ref{SLab2} shows that the phase transition will begin later as $\alpha$ increases for fixed $b$, while Fig. \ref{SLab4} tells us that the phase transition will take place at the same point for different $b$ if we fixed $\alpha$.

\section{conclusions}

The behaviors of the holographic superconductors/insulator
transition have been investigated by introducing a charged
scalar field coupled with a logarithmic electromagnetic field in
both the  Einstein-Gauss-Bonnet AdS black hole and soliton.

For the Einstein-Gauss-Bonnet AdS black hole,  we find that the
critical temperature increases as the value of $b$ increases with
fixed $\alpha$, which means that the larger parameter $b$ makes it
easier for the scalar hair to be condensate in the
Einstein-Gauss-Bonnet AdS black hole; however,  the critical
temperature decreases as $\alpha$ increases for the fixed $b$, which
means that the stronger Gauss-Bonnet coupling makes  condensated
harder in the black hole. We note that the gap frequency
$\omega_{g}$ decreases with the increase of the coupling parameter
$b$ for fixed  $\alpha$, but  it increases as $\alpha$ increases for
fixed $b$. The ratio of the gap frequency in conductivity $\omega_g$
to the critical temperature $T_c$ in the Einstein-Gauss-Bonnet AdS
black hole with the logarithmic electrodynamic field depends on both
the Gauss-Bonnet constant and the coupling parameter of logarithmic
electrodynamic field. We also show that the critical exponent is
independent of the parameters $b$ and $\alpha$, which is in
agreement with the value $1/2$. The result seems to be a universal
property for various nonlinear electrodynamics if the scalar field
$\psi$ takes the form of this paper.

For the Einstein-Gauss-Bonnet AdS Soliton, the system  is the
insulator phase when $\mu$ is small, but there is a phase transition
and the AdS soliton reaches the superconductor (or superfluid) phase
when $\mu$ larger than the critical chemical potential $\mu_{+ S}$
or $\mu_{- S}$. Especially, the phase transition can occur even at
strictly zero temperature. We note that the critical chemical
potential $\mu_{i S}$ ($i=\pm 1$) increases as $\alpha$ increases
with fixed $b$, while it does not change by the coupling constant
$b$ with fixed $\alpha$. That is to say, the phase transition will
begin later as $\alpha$ increases for fixed $b$, while the phase
transition will take place at the same point for different $b$ with
fixed $\alpha$.

\begin{acknowledgments}

This work was supported by the National Natural Science  Foundation
of China under Grant No. 11175065, 10935013; the National Basic
Research of China under Grant No. 2010CB833004; the SRFDP under Grant No. 20114306110003; PCSIRT, No. IRT0964; the Hunan Provincial Natural Science Foundation of China under Grant No 11JJ7001; and the Construct Program of the National
Key Discipline.

\end{acknowledgments}

\end{document}